\newcommand{\sbk}[1]{\left[ #1 \right]}

\newcommand{\MARU}[1]{{\ooalign{\hfil#1\/\hfil\crcr\raise .167ex\hbox{\mathhexbox20D}}}}

\documentclass[cpp]{w-art}
\usepackage{times}
\usepackage{w-thm}

\usepackage{graphicx,color}
\begin{document}
\DOIsuffix{theDOIsuffix}
\Volume{}
\Issue{}
\Month{}
\Year{2007}
\pagespan{1}{}
\Receiveddate{22 May 2007}



\title[MD simulation of  hydrogen isotope injection into graphene]{Molecular dynamics simulation of  hydrogen isotope injection into graphene}


\author[Hiroaki Nakamura]{Hiroaki Nakamura\footnote{Corresponding
     author: e-mail: {\sf hnakamura@nifs.ac.jp}, Phone: +81-572-58-2356,
     Fax: +81-572-58-2626,}\inst{1}} \address[\inst{1}]{National Institute for Fusion Science, 322-6 Oroshi-cho, Toki 509-5292, Japan}
\author[Arimichi Takayama]{Arimichi Takayama\footnote{takayama@nifs.ac.jp}\inst{1}}
\author[Atsushi Ito]{Atsushi Ito\footnote{ito.atsushi@nifs.ac.jp}\inst{1,2}}
\address[\inst{2}]{Department of Physics, Graduate School of Science, Nagoya University, Nagoya 464-8602, Japan}

\begin{abstract}
We reveal the hydrogen isotope effect of three chemical reactions, i.e, the reflection, the absorption and the penetration ratios, by classical molecular dynamics simulation with a modified Brenner's reactive empirical bond order (REBO) potential potential.
We find that the  reflection by $\pi-$electron does not depend on the mass of the incident isotope, but the peak of the  reflection by nuclear moves to higher side of incident energy.
In addition to the reflection, we also find that the absorption ratio in the positive $z$ side of the graphene becomes larger, as the mass of the incident isotope becomes larger.
On the other hand, the absorption ratio in the negative $z$ side of the graphene becomes smaller.
Last, it is found that the penetration ratio does not depend on the mass of the incident isotope because the graphene potential is not affected by the mass.
\end{abstract}
\maketitle                   

\renewcommand{\leftmark}{H. Nakamura et al.: MD simulation of  hydrogen isotope injection into graphene}

\section{Introduction}

Plasma-carbon interaction yields small hydrocarbon molecules on divertor region of a nuclear fusion device\cite{Nakano,Roth,Roth2,Mech,LHD}.
Diffusing from divertor region to  core plasma region of fusion device, generated hydrocarbon takes  energy from the core plasma.
Reduction of hydrocarbon diffusing from divertor is the main aim of studies in plasma-carbon research.
To achieve the aim,  researches with computer simulation  have been being done\cite{Alman,Ito,Ito2,Nakamura}.
However, the creation mechanism of the hydrocarbons has not been elucidated yet.

\begin{figure}[htb]
\includegraphics[width=.4\textwidth]{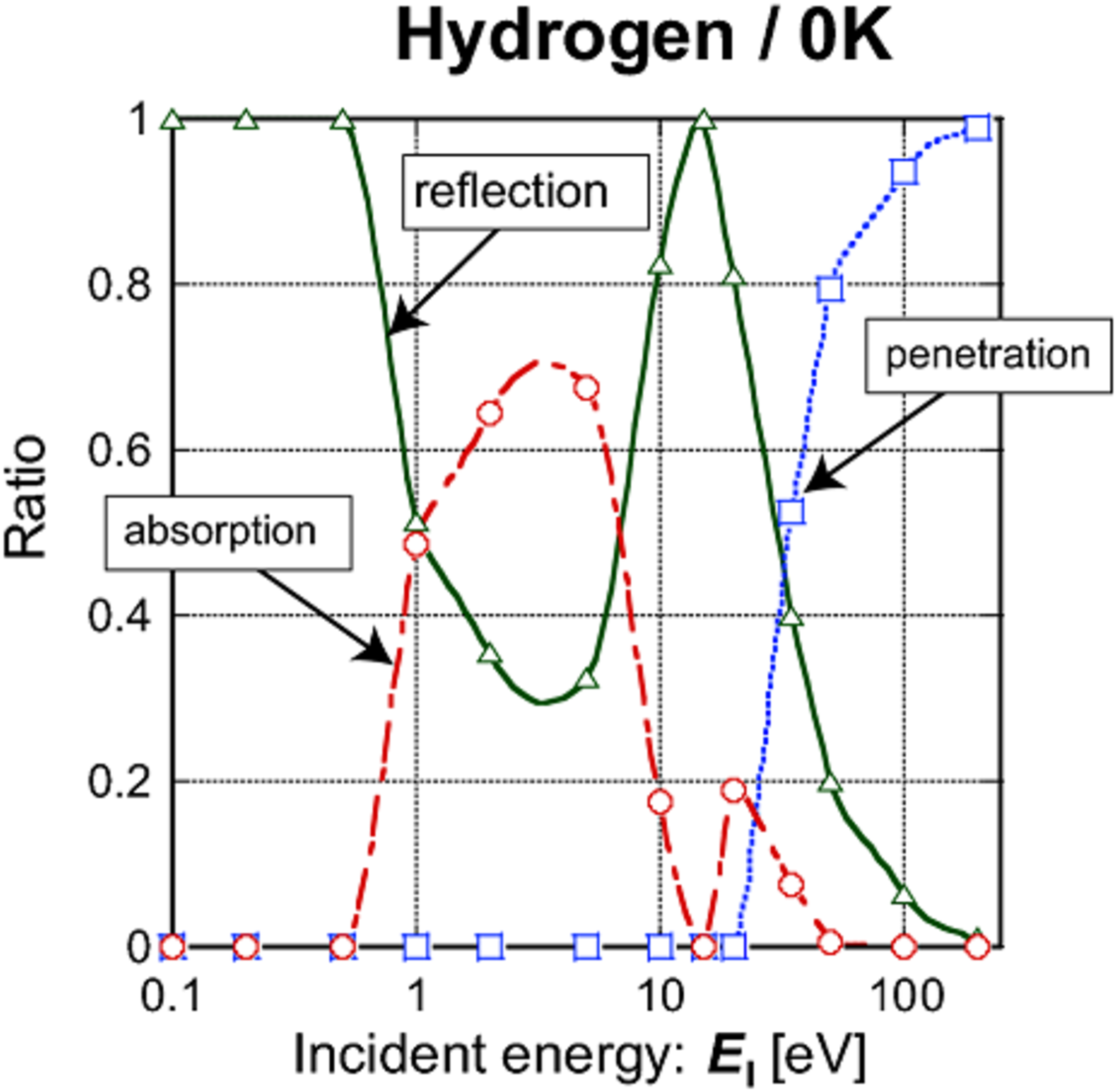}~a)
\hfil
\includegraphics[width=.4\textwidth]{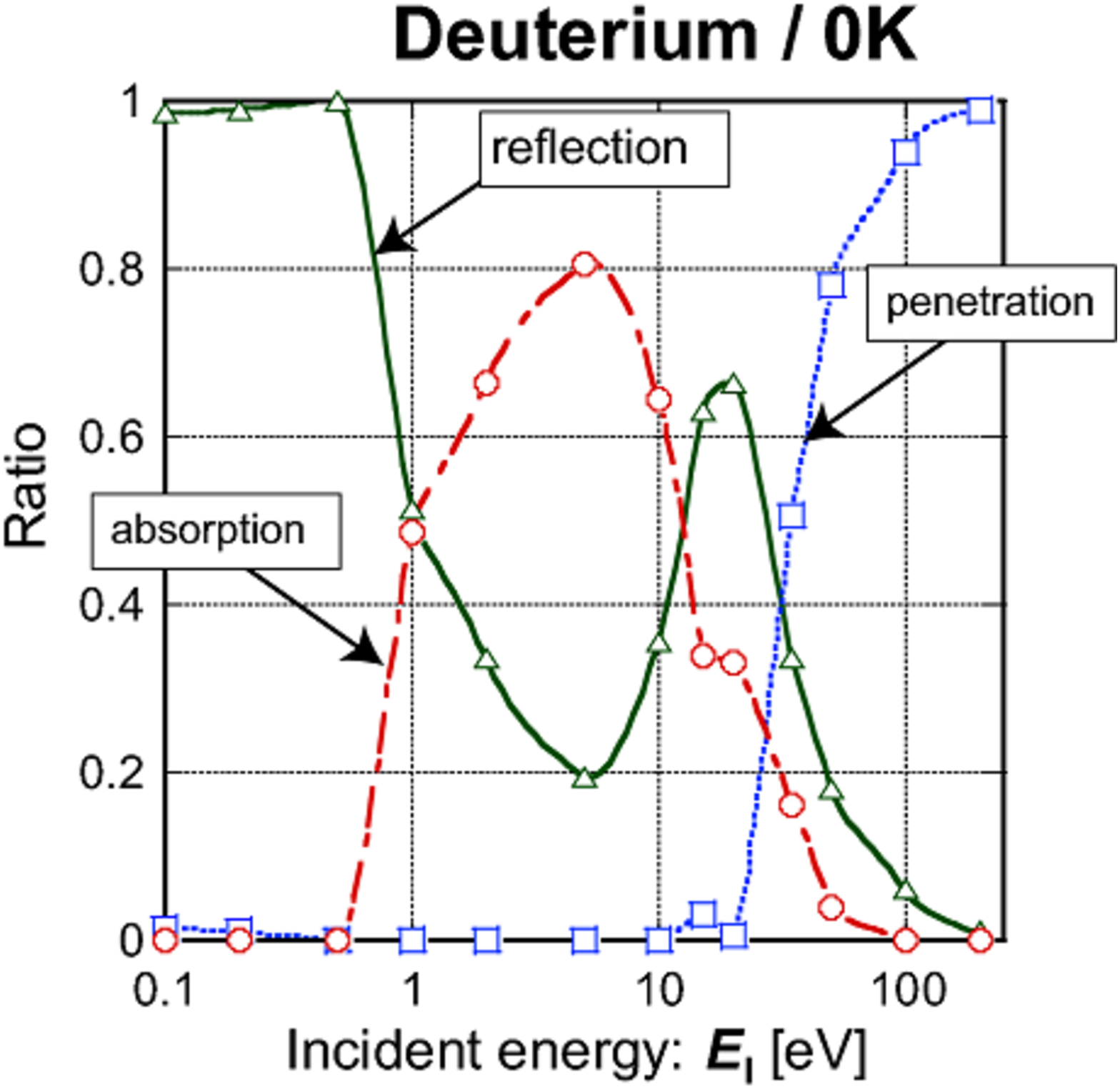}~b)
\\
\includegraphics[width=.4\textwidth]{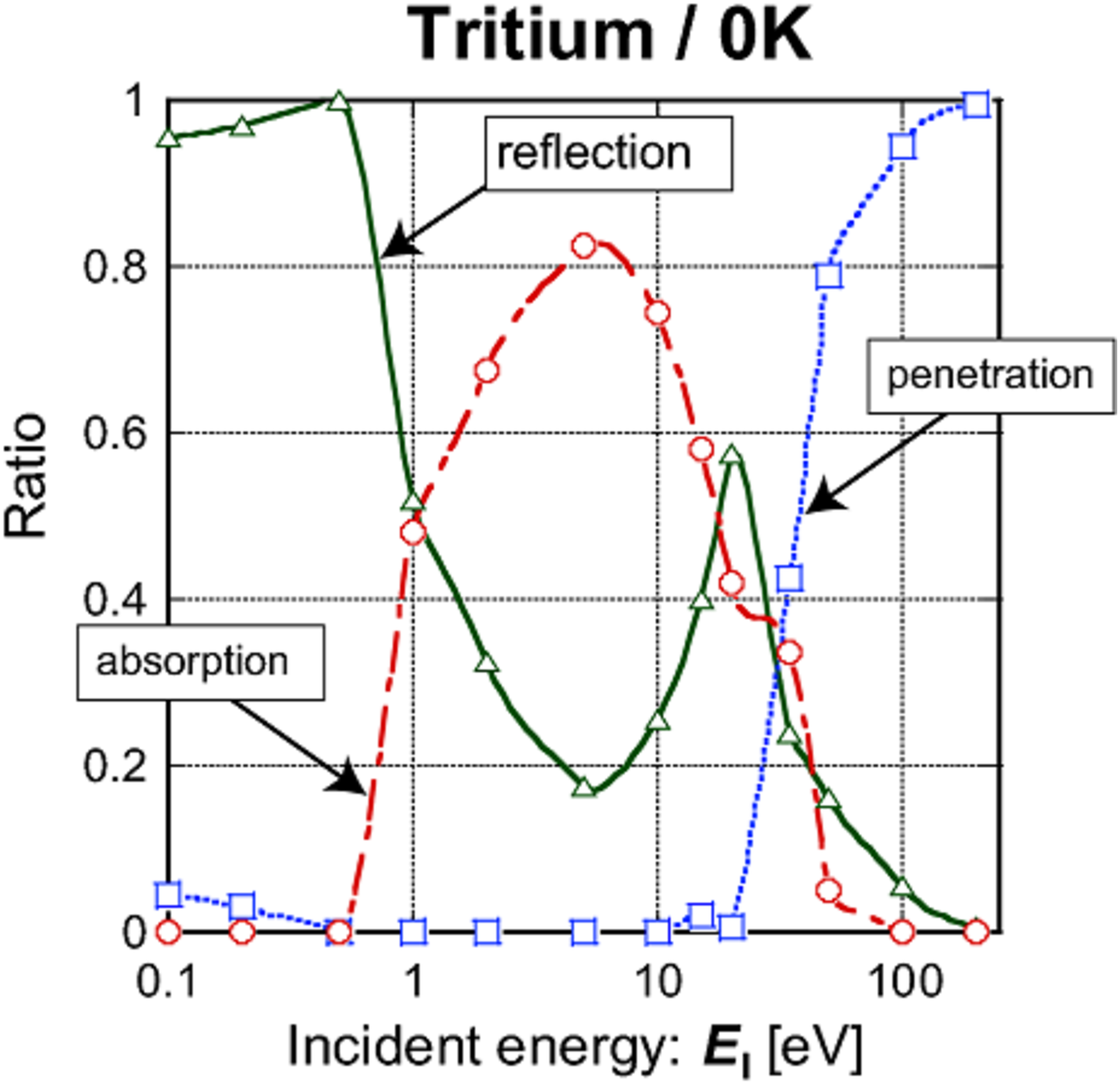}~c)
\caption{Incident energy dependence of the absorption, the reflection and the penetration ratios.
Three types of injected atoms are hydrogen (a), deuterium (b) and tritium (c). 
Open triangles, open circles, and open squares denote  denote the absorption, the reflection and the penetration ratios, respectively.
Dash-dotted lines, solid lines, and short-dashed lines are drawn as the guide for eyes for the absorption, the reflection and the penetration ratios, respectively.}
\label{fig:2}
\end{figure}

We, therefore, as the first step to clarify the creation mechanism, investigated, by computer simulation, collision process of  hydrogen atoms and one graphene sheet, which is regarded as one of   basic processes of  complex  plasma-carbon interaction in the previous works\cite{Ito,Ito2}.
From the previous works  in which an incident hydrogen kinetic energy $E_{\rm I}$ is less than 100 eV  to compare with experiments, it was found that an hydrogen-absorption ratio of one graphene sheet depends on the incident hydrogen energy,  and  that the collision mechanism between a graphene and a hydrogen can be  classified into three types of processes: absorption process,  reflection process,  and  penetration process (see Fig. \ref{fig:2}(a)).
Moreover, it was also found that when  hydrogen atom is absorbed by graphene,  the nearest carbon atom overhangs from the graphene which we called ``overhang structure".

Based on the above results, as the second step, simulation model were extended\cite{Nakamura} from a single  graphene sheet  to  multilayer graphene sheets, which is a more realistic sputtering process of graphene sheets and hydrogen atoms than the previous work\cite{Ito}.
From the second work\cite{Nakamura}, we found the following fact:
breaking the covalent bonds between carbon atoms by hydrogen  does not play an important role during destruction process of graphene structure, but  momentum transfer from incident hydrogen to graphene causes to destroy graphene structure.
Moreover, it was found\cite{Nakamura}, that almost all fragments of graphene sheets form chain-shaped molecules, and that  yielded hydrocarbon molecules are composed of carbon chain and single hydrogen-atom. 

In the present paper, we investigate hydrogen isotope effect for  collision process of  a single hydrogen isotope and one graphene sheet.
Information of dependence of the chemical reaction on a type of isotope is necessary to realize plasma confinement  nuclear fusion system.
In the present simulation, we change only the mass of the injected isotope, without changing the interaction potential.
We used  `classical' molecular dynamics (CMD) algorithm with  modified Brenner's reactive empirical bond order (REBO) potential which we proposed to deal with chemical reaction between hydrogen and graphene in the previous simulations\cite{Ito,Ito2,Nakamura,Brenner}.

\section{Simulation Method and Model}
We adapt CMD simulation with the \textit{NVE} condition, in which the number of particles, volume and total energy are conserved.
The second order symplectic integration\cite{Suzuki} is used to solve the time evolution of the equation of motion. 
The time step is $5 \times 10^{-18} \mathrm{~s}$.
The modified Brenner's reactive empirical bond order (REBO) potential\cite{Brenner} has the following form: 
\begin{eqnarray}
U \equiv \sum_{i,j>i} \sbk{V_{[ij]}^\mathrm{R}( r_{ij} )
 - \bar{b}_{ij}(\{r\},\{\theta^\mathrm{B}\},\{\theta^\mathrm{DH}\}) V_{[ij]}^\mathrm{A}(r_{ij}) }, \nonumber \\
	\label{eq:model_rebo}
\end{eqnarray}
where $r_{ij}$ is the distance between the $i$-th and the $j$-th atoms. 
The functions $V_{[ij]}^{\mathrm{R}}$ and $V_{[ij]}^{\mathrm{A}}$ represent repulsion and attraction, respectively.
The function $\bar{b}_{ij}$ generates multi--body force. (See details of the modified Brenner's REBO potential in Ref.\cite{Ito2}.)

In order to investigate the difference of the isotopes, i.e., hydrogen (H), deuterium (D) or tritium (T), we clarify the mass dependence of the injected isotope.
The value of the mass  for H, D, or T is shown in Table \ref{tab:4}.
The potential function is not changed for each isotope.

Simulation model is shown in Fig. \ref{fig:1}.
We inject the hydrogen isotope into the graphene composed of 160 carbon atoms.
The center of mass of the graphene is set to the origin of coordinates.
The surface of the graphene is parallel to the $x$--$y$ plane.
The size of the graphene is 2.13 nm $\times$ 1.97 nm.
The graphene has no lattice defects and no crystal edges due to periodic boundary condition toward $x$ and $y$ directions.
The structure of the graphene is used to the ideal lattice of graphene.
Each velocity of  carbon atoms of the graphene is zero in the initial state, that is, the initial temperature of the graphene is set to zero Kelvin.

The hydrogen isotope is injected parallel to the $z$ axis from $z = 4$~\AA.
We repeat 200 simulations where the $x$ and $y$ coordinates of injection points are set at random.
As a result, we obtain three chemical reaction ratios for H, D, or T by counting each a reaction.

\section{Results and Discussions}
We observed three kinds of reactions between the single hydrogen isotope atom and the graphene by CMD simulation, which are absorption reaction, reflection reaction and penetration reaction (see Fig. \ref{fig:2}).
We found the following differences  of the reflection and the absorption ratios among three isotopes.
On the other hand, the penetration ratio has almost the same $E_\textrm{I}$ dependence.

\subsection{Reflection  ratio}\label{sec.ref}
From the previous work\cite{Ito2}, it was found that two kinds of repulsive force work between the incident atom and the graphene.
One is derived by the $\pi-$electron over the graphene and the other is done by  nuclear of carbon.

\begin{figure}[htb]
\begin{minipage}{.45\textwidth}
\includegraphics[width=.9\textwidth]{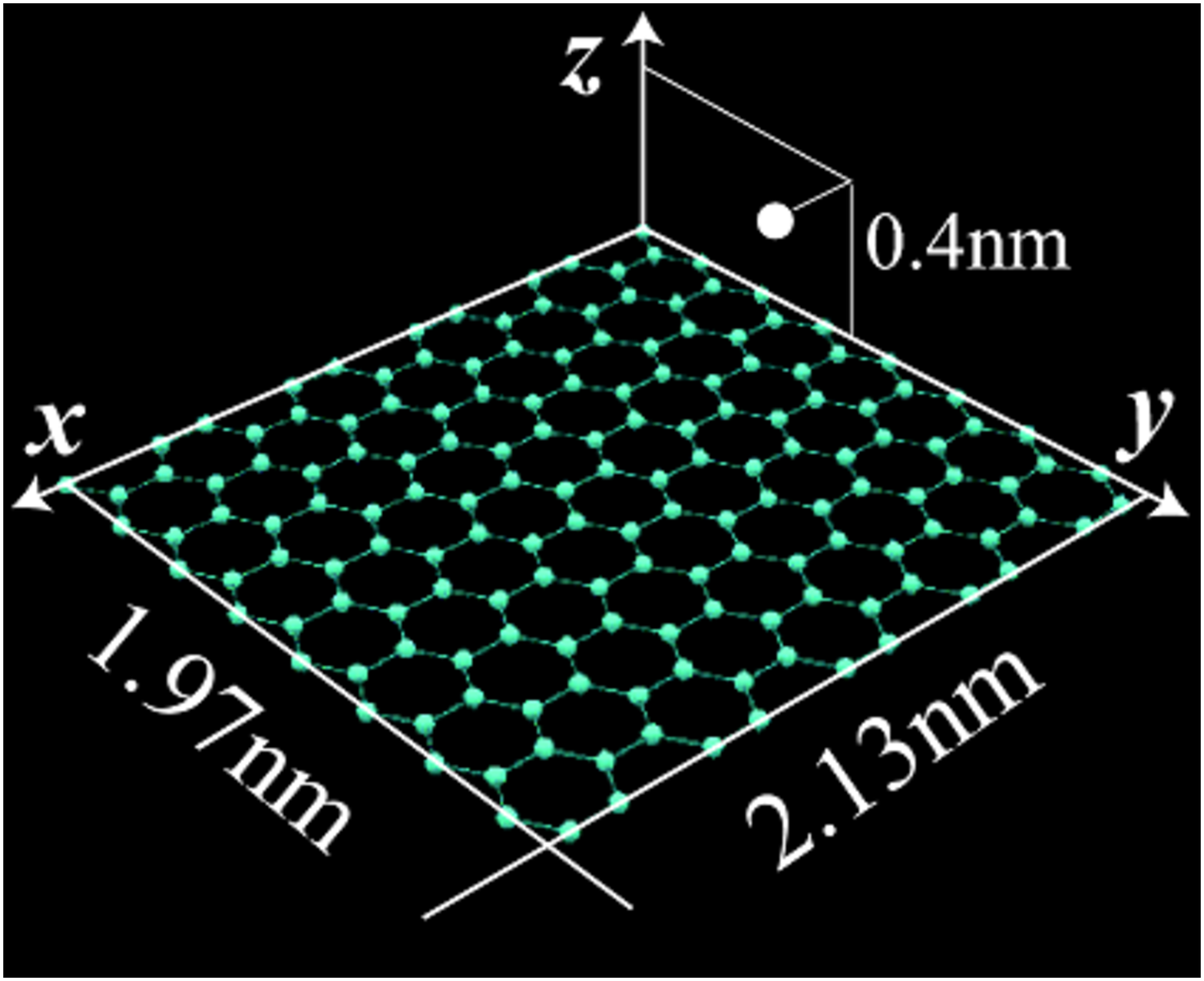}
\caption{Simulation model. There are 160 carbon atoms and an injected hydrogen isotope. 
The length of the covalent bond between carbons is 0.142nm at temperature $T=0$K.
The periodic boundary condition is used in $x$ and $y$ directions.
A hydrogen atom is injected parallel to the $z$ axis from $z=$0.4nm.
}
\label{fig:1}
\end{minipage}
\hfil
\begin{minipage}{.45\textwidth}
\includegraphics[width=.9\textwidth]{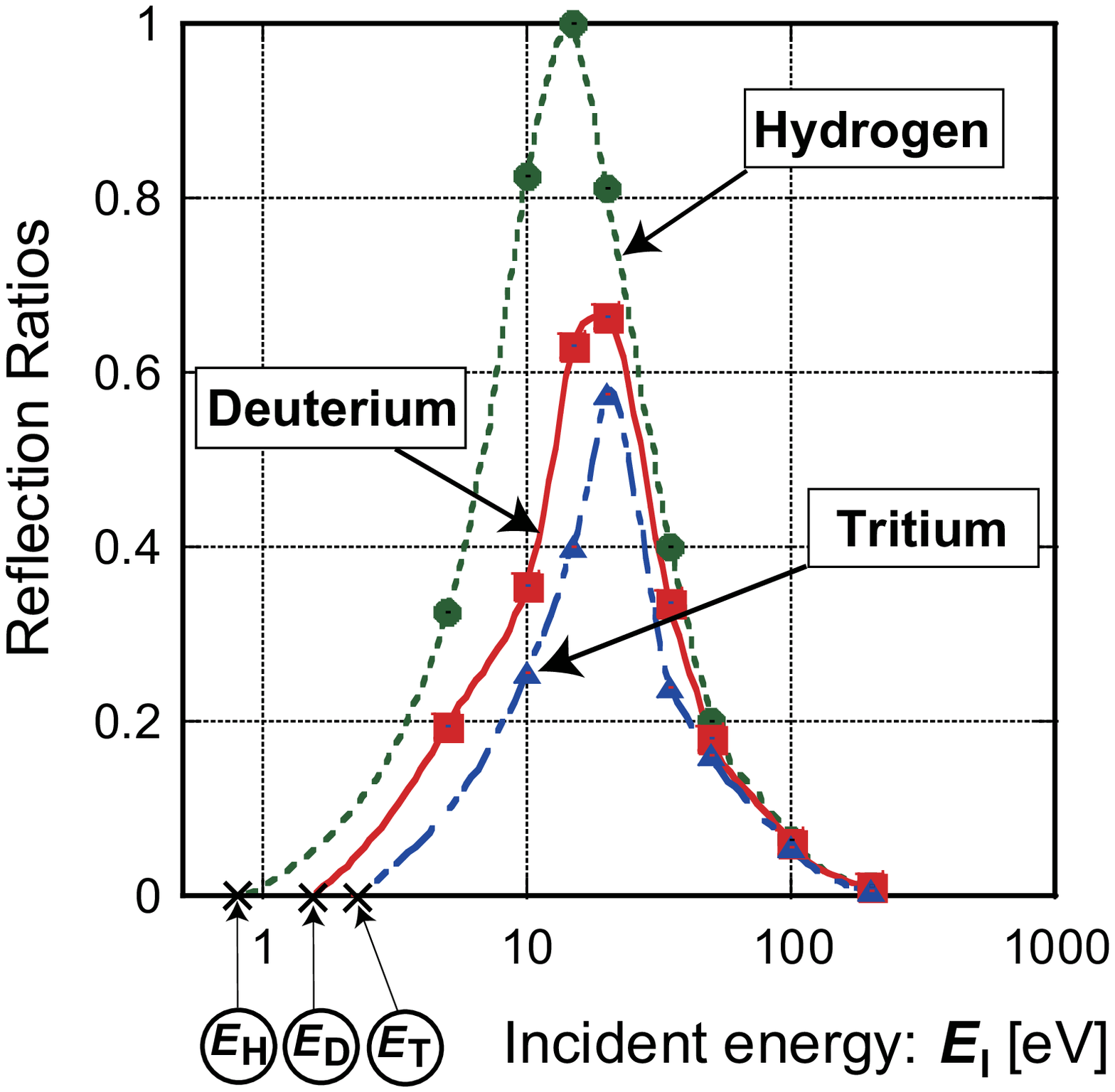}
\caption{The isotope mass $m$ dependence of the reflection ratio by the nuclear repulsion.
Filled circles, filled squares, and filled triangles denote the reflection ratios for hydrogen, deuterium and tritium, respectively.
These plotted data are picked up from Fig. \ref{fig:2}.  
The estimated minimum nuclear-reflection-energies $E_\textrm{H}, E_\textrm{D}$ and $E_\textrm{T}$ are given in Table \ref{tab:4}.
Dashed line, solid line and dash-dotted line are drawn as the guide for eyes.
}
\label{fig:ref}
\end{minipage}
\end{figure}

As the result of the present simulation,  the $E_\textrm{I}$ dependence has the following properties.
In the case of $E_\textrm{I} < 0.5 \textrm{ eV}$, the reflection ratio is almost one for all isotopes.
This behavior is  explained  by the fact that the reflection in this energy region is derived by the repulsive force of $\pi-$electrons over graphene surface\cite{Ito2}, which does not depend on the mass of the incident isotope.
As $E_\textrm{I}$ is getting larger than 0.5 eV, the reflection ratios are decreasing, and then it is increasing by the nuclear repulsive force of the carbon atom.
Around $E_\textrm{I} \sim 15$ eV, they have the peak (see Figs. \ref{fig:2} and \ref{fig:ref}).
Then, without the mass dependence, they decrease again  and approach to zero (Fig. \ref{fig:ref}), because the penetration reaction appears in the energy region that $E_\textrm{I} > 15 \textrm{ eV}.$
The details of the penetration reaction will appear in \S \ref{sec.pene}.

By comparison with three isotopes, it is found that the peak energy of the reflection ratio becomes larger as the mass is getting larger, but the  peak height becomes smaller (see Fig. \ref{fig:ref}).
This behavior can be explained by the reflection mechanism in the previous work\cite{Ito2}, where the incident energy $E_\textrm{I}$ has the following necessary condition for  the reflection reaction by the nuclear of carbon:
\begin{equation}
E_\mathrm{I} > E_\mathrm{ref}(m)\equiv 0.84\  \frac{m}{m_\mathrm{H} }\  \frac{m_\mathrm{C} + m_\mathrm{H}}{m_\mathrm{C} + m } \  \mathrm{[eV]},
\label{eq.ref}
\end{equation}
where the isotope mass $m$ dependence is modified from the previous equation\cite{Ito2} in order to cover  three isotopes. 
The hydrogen mass $m_\textrm{H}$ is 1.00794 u, and the carbon mass $m_\textrm{C}$ is 12.00000 u.
The value of $E_\textrm{ref}(m)$ is given in Table \ref{tab:4} and Fig. \ref{fig:ref} for all isotopes.

From Eq. (\ref{eq.ref}) or Table \ref{tab:4}, it is found that, as the mass of the incident atom becomes larger, it needs higher incident energy for the isotope to be reflected  by  the graphene.
The reflection reaction by the nuclear repulsion occurs at  $E_\textrm{I} \sim E_\textrm{ref}$, and then it increases monotonically until the penetration reaction becomes dominant among three reactions, as shown in Fig. \ref{fig:ref}
Therefore, in the case that the starting energy of the reflection reaction by the nuclear $E_\textrm{ref}$ is smaller, the reflection ratio can reach a higher value at the peak energy.
According to the above mechanism, the peak energy of the reflection ratio becomes larger and the peak height becomes smaller,  as the mass of the isotopes is getting heavier, as shown in Fig. \ref{fig:ref}.

\subsection{Absorption  ratio}
The absorption ratio has two peaks at $E_\textrm{I} \sim 5$eV and 24 eV. 
One peak denotes the overhang state in the positive $z$ side of the graphite, and the other peak is the overhang state in the negative $z$ side\cite{Ito2}.
From Fig. \ref{fig:2}, the height of the second peak, which is near 24 eV, becomes smaller, as the mass of the incident isotope is increasing.
On the other hand, the first peak of the absorption ratio, which is around 5 eV, becomes large.
The origin of the mass dependence of the absorption ratio is the same one as that of the reflection ratio in  \S \ref{sec.ref}.
The velocity becomes slower as the mass becomes large.
Therefore, it becomes easier for the graphene to trap the isotope in the positive $z$ side.

\subsection{Penetration  ratio}\label{sec.pene}
From Fig. \ref{fig:2}, it is found that the penetration ratio does not depend on the mass of the incident isotope, i.e., H, D and T.
The incident atom must overcome the graphene potential to penetrate the graphene sheet.
The graphene potential does not depend on the mass of the incident atom.
Therefore, the penetration ratio does not depend on the mass of the isotope.


\section{Summary}
We reveal the hydrogen isotope effect of three chemical reactions, i.e, the reflection, the absorption and the penetration reactions, by CMD simulation with the modified Brenner's REBO potential.
From the previous work\cite{Ito2}, the reflection process is divided into two processes, that is, reflection by $\pi-$electron and by nuclear.
In the present work, we find that the  reflection by $\pi-$electron does not depend on the mass of the incident isotope, but the peak of the  reflection by nuclear moves to higher side of $E_\textrm{I}$.
In addition to the reflection, we also find that the absorption ratio in the positive $z$ side of the graphene becomes larger, as the mass of the incident isotope becomes larger.
On the other hand, the absorption ratio in the negative $z$ side of the graphene becomes smaller.
Last, it is found that the penetration ratio does not depend on the mass of the incident isotope because the graphene potential is not affected by the mass.

\begin{table}[htb]
\caption{The mass and the reflection threshold energy by nulcear repulsive force $E_\mathrm{ref}$ for hydrogen, deuterium and tritium.
The reflection threshold energy $E_\mathrm{ref}$ depends on the mass of the isotope as Eq. (\ref{eq.ref}). 
We use the unified atomic mass u as   the unit of mass.}
\label{tab:4}\renewcommand{\arraystretch}{1.5}
\begin{tabular}{cccc} \hline
& Hydrogen & Deuterium & Tritium \\ \hline
$m$ [u] &  1.00794  $( \equiv m_\mathrm{H}) $ & 2.01410$( \equiv m_\mathrm{D}) $ & 3.01605$( \equiv m_\mathrm{T}) $ \\
$E_\mathrm{ref} (m)$ [eV] & 0.84$( \equiv E_\mathrm{H}) $ & 1.56$( \equiv E_\mathrm{D}) $ & 2.18$( \equiv E_\mathrm{T}) $ \\ \hline
\end{tabular}
\end{table}

\begin{acknowledgement}
The authors acknowledge Dr. Noriyasu Ohno for helpful comments.
Numerical simulations are carried out by use of the Plasma Simulator at National Institute for Fusion Science.
The work is supported partly by Grand-in Aid for Exploratory Research (C), 2006, No.~17540384 from the Ministry of Education, Culture, Sports, Science and Technology and  partly by the National Institutes of Natural Sciences undertaking for Forming Bases for Interdisciplinary and International Research through Cooperation Across Fields of Study, and Collaborative Research Programs (No. NIFS06KDAT012, No. NIFS06KTAT029, No. NIFS07USNN002 and No. NIFS07KEIN0091).
\end{acknowledgement}

\end{document}